\documentclass[10pt,letterpaper,twocolumn]{article} 

\usepackage{ol2}
\usepackage[draft,implicit=false]{hyperref}
\usepackage{amsmath}

\begin{document}

\twocolumn[ 

\title{Positive Operator-Valued Measure reconstruction of a beam-splitter tree based photon-number-resolving detector}


\author{F. Piacentini,$^{1,*}$ M. P. Levi,$^{1,2}$ A. Avella,$^1$ M. Lopez,$^3$ S. K\"{u}ck,$^3$, S. V. Polyakov,$^4$ I. P. Degiovanni,$^1$ G. Brida,$^1$ and M. Genovese$^{1,5}$}

\address{
$^1$Istituto Nazionale di Ricerca Metrologica (INRIM),  Strada delle Cacce 91, 10135 Torino, Italy.
\\
$^2$Dipartimento di Fisica, Politecnico di Torino,  Corso Duca degli Abruzzi 24, Torino 10129, Italy.
\\
$^3$Physikalisch-Technische Bundesanstalt Braunschweig,  Bundesallee 100, 38116 Braunschweig, Germany.
\\
$^4$National Institute of Standard and Technology,  100 Bureau Dr. Stop 8410, Gaithersburg, MD 20899, U.S.
\\
$^5$Consorzio Nazionale Interuniversitario per le Scienze Fisiche della Materia (unit\`{a} di Torino Universit\`{a}),  via P. Giuria 1, 10125 Torino, Italy
\\
$^*$Corresponding author: f.piacentini@inrim.it
}

\begin{abstract}
Here we present a reconstruction of the Positive Operator-Value Measurement of a photon-number-resolving detector comprised of three 50:50 beamsplitters in a tree configuration, terminated with four single-photon avalanche detectors. The four detectors' outputs are processed by an electronic board that discriminates detected photon number states from 0 to 4 and implements a ``smart counting'' routine to compensate for dead time issues at high count rates.
\end{abstract}


 ] 

Photon-number-resolving (PNR) detectors \cite{had09, migdallBook}, i.e. photodetectors that can resolve the number of photons that are impinging on them, have achieved a critical role in a wide variety of research fields, ranging from quantum mechanics foundations experiments \cite{Gen05} to quantum metrology \cite{pol09,zwi10}, imaging \cite{Bra08,Brida2009} and information \cite{gis07,ben10}.
As a consequence, a precise quantum characterization of these devices has become crucial \cite{Gen05,pol09,zwi10,Bra08,Brida2009,gis07,ben10,brie2010,b05}.
In a quantum mechanical framework, a full operational description of a PNR device is its \textit{positive operator-valued measure} (POVM), i.e. the set of operators $\widehat{\Xi}_n$ describing a physical process that leads to a particular measurement outcome $n$.
A measurement of the elements of a detector's POVM can be quite non-trivial, because one has to carefully choose the best-suited technique for a tomographic reconstruction of the POVM of the device under test, depending on its particular properties  \cite{Dar09,Lun09,Bri11,Mog10,zhang2012,zhang2012a,PRL_POVM}.

There exist different types of PNR detectors, e.g. photo-multiplier tubes \cite{burle,burle1}, hybrid photo-detectors \cite{NIST,all14}, quantum-dot field-effect transistors \cite{gan07}, multipixel counters \cite{kala2011}, visible light photon counters \cite{VLPC1,VLPC2,VLPC3}, and superconducting Transition Edge Sensors (TESs) \cite{fuk11,pre09,cab08,lit08,lol11,lol13,ros05,bandler06,irw05,Nam14,Pfi14}.
Some of those detector families hold a significant promise for future applications, even if their use at present is very difficult because of a large experimental overhead associated with their operation.
On the other hand, even though traditional single-photon avalanche detectors (SPADs) are only capable to discriminate between zero and one (or more) detected photons, photon number resolution can be obtained by multiplexing those detectors spatially \cite{multiSpatial,multiSpatial1} or temporally \cite{multiTemporal1,multiTemporal2,multiTemporal3,multiTemporal4}.
At present, this solution is by far the easiest and cheapest way to achieve a photon number resolving capability, even though at a cost of sacrificing linearity due to detector saturation \cite{rad14}. 
Here we present the POVM reconstruction of a multiplexed PNR detector (at 1550 nm) composed of four Indium/Gallium arsenide (InGaAs) SPADs connected to a beam-splitter (BS) tree made with three 50:50 fiber BSs. The output of the InGaAs SPADs is processed with a field-programmable gate array (FPGA) board, giving as output the detected photon number (up to 4 detected photons per pulse).

Because this detector is not phase-sensitive, its POVM is diagonal in the Fock states basis:
\begin{equation}\label{POVM}
\widehat{\Xi}_n=\sum_m \Xi_{nm} |m\rangle \langle m| \;\;\;\; \left( \sum_n \widehat{\Xi}_n = \mathbf{I} \right)
\end{equation}
where the $\Xi_{nm}=\langle m| \widehat{\Xi}_n| m\rangle$ elements give the detector tree probability of counting $n=0,...,4$ photons with $m$ impinging photons per pulse.
To reconstruct $\Xi_{nm}$, we test the response of our device to a set of $J$ coherent states. The response of our PNR detector to the $j$-th coherent state input $|\alpha_j\rangle$ can be written as:
\begin{equation}\label{eq:stat_model}
\xi_{nj} =\hbox{Tr}\left[|\alpha_j\rangle\langle\alpha_j| \widehat{\Xi}_n\right] =\sum_m \Xi_{nm}\, a_{mj}
\end{equation}
where $a_{mj}=\exp(-\left|\alpha_j\right|^2)\left|\alpha_j\right|^{2m}/m!$ gives the probability that there are exactly $m$ photons in one pulse sampeled from a coherent state $j$ (i.e. with the mean photon number $|\alpha_j |^2$).

\begin{figure}[htb]
\centerline{\includegraphics[width=9cm]{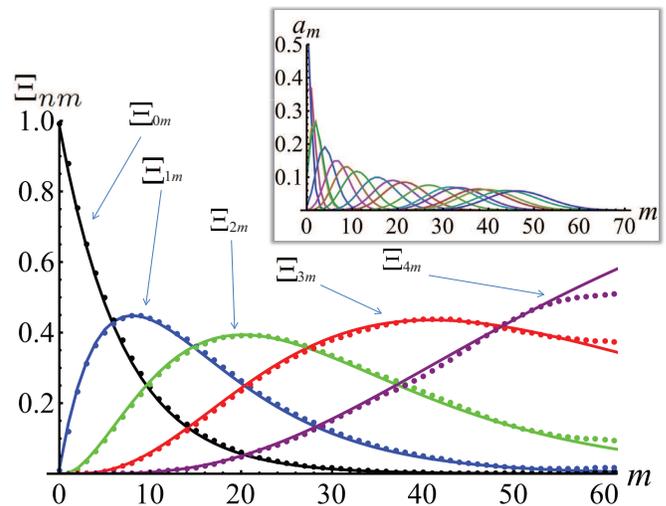}}
\caption{Main plot: reconstruction of the POVM elements of a detector tree. The solid lines (black for $n=0$, blue for $n=1$, green for $n=2$, 
red for $n=3$, purple for $n=4$) represent the theoretical POVM, while the corresponding dots are the reconstructed POVM elements.
Inset: photon number distributions (up to $m=70$ incoming photons) of the $J$ coherent probes used in this POVM reconstruction.}
\label{POVMrec}
\end{figure}

\begin{figure}[htb]
\centerline{\includegraphics[width=8.5cm]{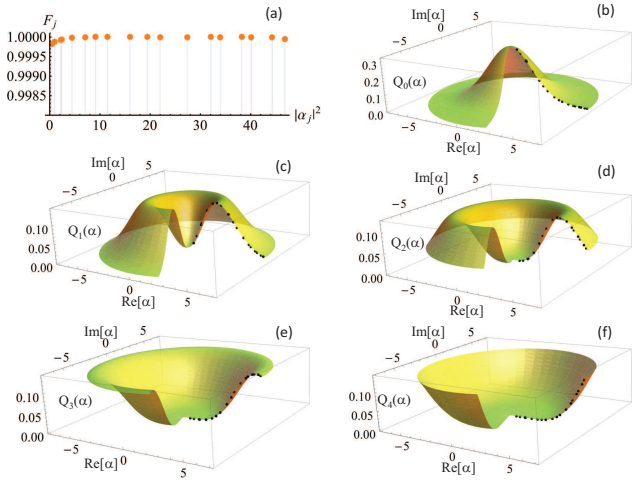}}
\caption{(a): Fidelities of the reconstructed probabilities of detecting $n$ photons ($\xi_{nj}^{(r)}$) versus experimentally obtained probabilities ($\xi_{nj}^{(e)}$), for each of the probe states $|\alpha_j\rangle$ (see eq. (\ref{fid})).
(b)-(f): Q-representation functions $\mathcal{Q}_n(\alpha)$ (labels ``b'',...,``f'' correspond to $n=0,...,4$) obtained exploiting the reconstructed POVM elements. The superimposed black dots represent the experimentally measured corresponding quantities $\pi^{-1} \xi_{nj}$.}
\label{F_j}
\end{figure}

Once we have measured the different $\xi_{nj}$ probabilities experimentally, we reconstruct
the $\Xi_{nm}$ elements by minimizing the quantity:
\begin{equation}\label{LeastS}
\sum_{nj} \left(\sum_{m=0}^{\infty} a_{mj}\, \Xi_{nm} - \xi_{nj}\right)^2,
\end{equation}
with the additional constraints of normalization ($\sum_{n=0}^{4} \Xi_{nm}=1,\;\forall~m$) and a ``smoothness'' condition given by a convex, quadratic and device-independent function regularizing the fluctuations of the reconstructed POVM elements \cite{Lun09,Bri11}.
There is no upper limit on the number of photons per pulse for a coherent state. However, it is impractical to consider an infinite space of impinging Fock states. Therefore we should restrict ourselves to a carefully chosen finite subspace. In particular, we perform the reconstruction over a finite space truncated at a certain value $M$ for which the probability of having $m\geq M$ photons in the brightest state is negligible within the accuracy of the reconstruction.
The inset of Fig. \ref{POVMrec} shows 18 probability distributions $a_{mj}$. Note that for each $m$ up to $m \simeq 50$, there are at least four probability distributions $a_{mj}$ that are differ from zero significantly. This is important to provide enough input for a meaningful minimization of the quantity in Eq. (\ref{LeastS}).
\begin{figure}[htb]
\centerline{\includegraphics[width=8.cm]{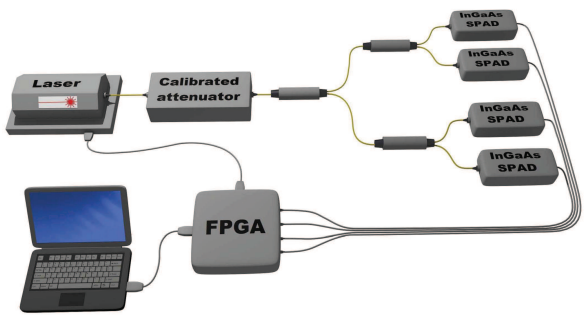}}
\caption{Experimental setup: a 90 kHz pulsed laser ($\lambda=1550$ nm) is sent to a calibrated attenuator whose output goes into a detector tree, comprised of four SPADs connected to a cascade of three 50:50 fiber BS. Outputs of the detectors, together with the laser sync signal, are sent to an FPGA, responsible for SPADs gating and real-time data processing.}
\label{setup_det-tree}
\end{figure}
The experimental setup (Fig. \ref{setup_det-tree}) is comprised of a 1550~nm pulsed laser, and an attenuator, here a half wave plate and a polarizing beam-splitter. The laser beam is fiber-coupled and sent onto the PNR detector to be characterized.
Our PNR detector is comprised of four InGaAs-SPADs connected to a BS-tree. The electrical output of the detectors is sent to an FPGA, that outputs a measured photon number between 0 and 4 in real-time \cite{gold13}. The same FPGA is used for gating the SPADs.
Note that, in general, the dead-time significantly affects the measurement accuracy. One way to avoid the dead-time is to choose a low repetition rate to guarantee that all the detectors will be operational for the next incoming pulse. This reduces data acquisition rate very significantly.
Instead, we implemented a different approach: the FPGA control circuit monitors the timing of the photo-electronic detections and only allows gating the SPADs from the laser when all the detectors are ready to count.
Even though dead-time avoidance may result in a somewhat lower data acquisition rate (with respect to the source emission rate), the full, unsaturated state of the detector is guaranteed for each recorded detection and provides an advantage over selecting a really low repetition rate to avoid dead-time issues altogether.
The experiment consists of probing our PNR detector with $J=18$ different coherent states with $|\alpha_j|^2$ ranging from $0.5$ to $46.8$ photons per pulse, generated by a pulsed laser with a repetition rate of 90 kHz.
After data acquisition, the four SPADs comprising the detector tree have been properly calibrated with a detector substitution technique \cite{but98}, i.e. by comparing the SPADs response with a calibrated power meter, and using a CW fiber laser at 1550 nm passing through a calibrated attenuator as a source.
In order to achieve a better accuracy, we calibrate our device as a whole, without disconnecting the SPADs from the beam-splitter tree, thus attributing the BS tree losses and asymmetric splitting to the overall ``detection efficiency'' of the detector at the end of each of the four branches of the detector tree: the four values obtained are $\eta_a=(12.70\pm0.07)\%$, $\eta_b=(13.75\pm0.08)\%$, $\eta_c=(14.10\pm0.07)\%$ and $\eta_d=(12.7\pm0.1)\%$.
Further, we estimated the probability of dark-click per gate for each detector, and obtained $p_{a,dark}=(1.20\pm0.03)\times10^{-4}$, $p_{b,dark}=(1.25\pm0.03)\times10^{-4}$, $p_{c,dark}=(1.13\pm0.01)\times10^{-4}$  and $p_{d,dark}=(2.52\pm0.02)\times10^{-4}$. The probability of an afterpulse per gate is negligible due to the long deadtime selected.\\
The reconstruction of the $\Xi_{nm}$ elements (dotted lines) up to 60 incoming photons are presented in Fig. \ref{POVMrec},
the main result of this paper.
The solid curves show the behavior of the theoretical POVM of our PNR device. The complete expression is too lengthy to be shown here \cite{migdallBook}, but that can be easily derived starting from the functional
$
g[x_a, x_b, x_c, x_d]= \sum_{  [i,j,k]_n }   \frac{ n! x_a [i] x_b [j] x_c [k] x_d [n-i-j-k] }{4^n  i! j! k! (n-i-j-k)!}
$
where the sum indexes $i$, $j$, $k$ go from 0 to the upper bound given by $i+j+k \leq n$.
The $x_{\gamma} [k]$ can be either the ``no-click'' probability of the $\gamma$-th detector in the presence of $k$ photons (referred to as $\mathcal{N}_{\gamma} [k]= (1-\eta_{\gamma})^k (1-p_{\gamma,dark}) $), or the corresponding ``click'' probability ($\mathcal{C}_{\gamma} [k]=1-\mathcal{N}_{\gamma} [k]$), where $\eta_{\gamma}$ is the detection efficiency of the $\gamma$-th detector, and $p_{\gamma,dark}$ is a probability of a ``click'' due to dark-counts per pulse.
Thus, the POVM coefficients $\Xi_{nm}$ can be evaluated as sum of the functional $g$ with the appropriate permutation of the elements of the vector $\mathbf{x}=(x_a, x_b, x_c, x_d)$ corresponding to $m$ clicks of the detector tree, i.e. $\Xi_{nm}= \sum_{\mathbf{x}=\{ m \mathcal{C}, (n-m) \mathcal{N}\} } g[\mathbf{x}]$ \cite{footnote}.\\
Fig. \ref{POVMrec} shows an excellent agreement between our reconstruction and the theoretical expectations up to $m\simeq 50 $ incoming photons per pulse.
The faithfulness of the reconstructed $\Xi_{nm}$ values rapidly decreases for $m > 50$. This happens due to insufficient statistics for higher-order photon-number states. In fact, it can be observed that with the set of probe states used, for $m>50$ the probability to generate such bright states rapidly decreases, so that the reconstruction algorithm that minimizes Eq. (\ref{LeastS}) suffers due to insufficient data.\\
To test the quality of our reconstructed POVM, we calculate reconstruction fidelity for each coherent state $|\alpha_j\rangle$:
\begin{equation}\label{fid}
F_j=\sum_{n=0}^4\sqrt{\xi_{nj}^{(e)}\cdot \xi_{nj}^{(r)}}
\end{equation}
where $\xi_{nj}^{(e)}$ is the measured probability of detecting $n$ photons with our PNR detector, given the $j$-th probe input, and $\xi_{nj}^{(r)}$ is the corresponding value obtained substituting the reconstructed POVM elements in Eq. (\ref{eq:stat_model}).
All the fidelity values are above $99.98\%$, as presented in Fig. \ref{F_j} (a). Such a high fidelity demonstrates the robustness and reliability of our method.\\
The $\xi_{nj}$'s are also strictly related to the Husimi $(Q-)$ representation of the POVM elements $\widehat{\Xi}_n$.\\
The $Q-$representation of the POVM element $\widehat{\Xi}_n$ is defined as $\mathcal{Q}_n(\alpha)= \pi^{-1} \langle \alpha | \widehat{\Xi}_n  | \alpha \rangle $ and, because of the phase independence of the $\widehat{\Xi}_n$'s (see Eq. (\ref{POVM})), the function $\mathcal{Q}_n(\alpha)$ is invariant to rotations with respect to the origin of the phase-space.
We plotted $\mathcal{Q}_n(\alpha)$ in the phase space in Fig. \ref{F_j} (b)-(f) for $n=0,...,4$, respectively.
obtained from the reconstructed POVM elements.
They are compared with experimentally measured values of $\mathcal{Q}_n(\alpha)$, i.e. $\pi^{-1} \xi_{nj}$, represented by the black dots. Without any loss of generality, because of the phase-invariance property of the $\mathcal{Q}_n(\alpha)$, these experimental values are arbitrarily placed at phase zero (i.e. $ \alpha_j=| \alpha_j | $).
The excellent match of the dots with the ``reconstructed'' $\mathcal{Q}_n(\alpha)$ is yet another demonstration of the quality of our reconstruction. It further highlights that the lack of faithfulness of the $\Xi_{nm}$ for $m \geq 50$, is consistent with the lack of statistical experimental data for the brightest photon-number states, i.e. it does not affect the accuracy of measuring $\mathcal{Q}_n(\alpha)$ and  $\xi_{nj}^{(r)}$.\\
In conclusion, we have reconstructed the POVM of a photon-number-resolving detector based on a tree of single-mode fiber beam-splitters connected to four InGaAs SPADs, i.e. with a counting capability up to four photons. We show a faithful reconstruction of the POVM up to 50 incoming photons per pulse, with fidelities $>99.98\%$.\\
Our setup features an FPGA-based active-control of the calibration apparatus that prevents issues associated with the dead-time of the SPADs, allowing a measurement only when all of the detectors are ready to count.\\
Because this PNR detector is made of off-the-shelf components, we believe that our technology is ready for an immediate, widespread use in the quantum information framework and related research fields.\\

This effort was supported by: EU FP7 under grant agreement n. 308803 (BRISQ2), by a project EXL02 - SIQUTE of the EMRP (The EMRP is jointly funded by the EMRP participating 
countries within EURAMET and the European Union), by MIUR (FIRB Grants No. RBFR10UAUV and No. RBFR10VZUG), by NATO (Grant No. 984397) and by Compagnia di San Paolo.
IPD thanks Matteo Paris for useful discussions.


\begin{thebibliography}{99}


\bibitem{had09} R.H. Hadfield, \emph{Nat. Phot.} \textbf{3},  696 (2009).

\bibitem{migdallBook} A. Migdall, S. V. Polyakov, J. Fan, and J. C. Bienfang Ed.s, \textit{Single-Photon Generation and Detection: Physics and Applications} (Academic Press; 2013)

\bibitem{Gen05} M. Genovese, \textit{Phys. Rep.} {\bf 413}, 319 (2005) and ref.s therein.

\bibitem{pol09} S.V. Polyakov and A. Migdall, \textit{J. Mod. Opt.} \textbf{56}, 1045 (2009).

\bibitem{zwi10} J. Zwinkels, E. Ikonen, N.P. Fox, G. Ulm and M.L. Rastello, \textit{Metrologia} \textbf{47}, R15 (2010).

\bibitem{Brida2009} G. Brida, L. Caspani, A. Gatti, M. Genovese, A. Meda and I. Ruo Berchera, \textit{Phys. Rev. Lett.} \textbf{102}, 213602 (2009).

\bibitem{Bra08} G. Brida, M. Genovese and I. Ruo Berchera, \textit{Nat. Phot.} {\bf 4}, 227 (2010).



\bibitem{gis07} N. Gisin and R. Thew, \textit{Nat. Phot.} \textbf{1}, 165 (2007).

\bibitem{ben10} \textit{Quantum Information, Computation and Cryptography}, Lect. Notes Phys. {\bf 808}, F. Benatti, M. Fannes, R. Floreanini and D. Petritis (Eds), (Springer, Berlin, 2010).
















\bibitem{brie2010} G. Brida, I.P. Degiovanni, A. Florio, M. Genovese, P. Giorda, A. Meda, M.G.A. Paris and A. Shurupov \textit{Phys. Rev. Lett.} \textbf{104}, 100501 (2010).

\bibitem{b05} G.Brida et al., \textit{J. Opt. Soc. Am. B} \textbf{22}, 488 (2005).







\bibitem{Dar09} G. D'Ariano, L. Maccone and P. Lo Presti \textit{Phys. Rev. Lett.} {\bf 93}, 250407 (2004).

\bibitem{Lun09} J.S. Lundeen, A. Feito, H. Coldenstrodt-Ronge, K.L. Pregnell, C. Silberhorn, T.C. Ralph, J. Eisert, M.B. Plenio and I.A. Walmsley, \textit{Nat. Phys.} \textbf{5}, 27 (2009).

\bibitem{Bri11} G. Brida, L. Ciavarella, I.P. Degiovanni, M. Genovese, L. Lolli, M.G. Mingolla, F. Piacentini, M. Rajteri, E. Taralli and M.G.A. Paris, \textit{New J. Phys.} \textbf{14}, 085001 (2012).

\bibitem{Mog10} D. Mogilevtsev,\textit{Phys. Rev. A} {\bf 82}, 021807(R) (2010).

\bibitem{zhang2012}  L. Zhang, H.B. Coldenstrodt-Ronge,	A. Datta, G. Puentes, J.S. Lundeen,	X. Jin,	B.J. Smith,	M.B. Plenio and I.A. Walmsley, \textit{Nat. Phot.} \textbf{6}, pp. 364-368 (2012).

\bibitem{zhang2012a} L. Zhang, A. Datta, H.B. Coldenstrodt-Ronge, X. Jin, J. Eisert, M.B. Plenio and I.A. Walmsley, \textit{New J. Phis.} 14, 115005 (2012).

\bibitem{PRL_POVM} G. Brida, L. Ciavarella, I.P. Degiovanni, M. Genovese, A. Migdall, M.G. Mingolla, M.G.A. Paris, F. Piacentini, and S.V. Polyakov, \textit{Phys. Rev. Lett.} \textbf{108}, 253601 (2012).

\bibitem{burle} G. Zambra, M. Bondani, A.S. Spinelli and A. Andreoni, \textit{Rev. Sci. Instrum.} {\bf 75}, 2762 (2004).

\bibitem{burle1} G.Q. Zhang, X.J. Zhai, C.J. Zhu, H.C. Liu and Y.T. Zhang, \textit{Int. J. of Quant. Inf.} \textbf{10}, 1230002 (2012).

\bibitem{NIST} M. Ramilli, A. Allevi, A. Chmill, M. Bondani, M. Caccia and A. Andreoni, \textit{J. Opt. Soc. Am. B} \textbf{27}, 852 (2010).

\bibitem{all14} A. Allevi and M. Bondani, \textit{J. Opt. Soc. Am. B} \textbf{31}, B14 (2014).

\bibitem{gan07} E.J. Gansen, M.A. Rowe, M.B. Greene, D. Rosenberg, T.E. Harvey, M.Y. Su, R.H. Hadfield, S.W. Nam and R.P. Mirin, \textit{Nat. Phot.} {\bf 1}, 585 (2007).


\bibitem{kala2011} D. A. Kalashnikov, S-H Tan, M. V. Chekhova, L. Krivitsky, Opt. Express, \textbf{19} 9352 (2011).

\bibitem{VLPC1} S. Takeuchi, J. Kim, Y. Yamamoto, and H. H. Hogue, Appl. Phys. Lett. \textbf{74}, 1063 (1999).

\bibitem{VLPC2} J. Kim, Y. Yamamoto, and H. H. Hogue, Appl. Phys. Lett. \textbf{70}, 2852 (1997).

\bibitem{VLPC3} E. Waks, K. Inoue, W. D. Oliver, E. Diamanti, and Y. Yamamoto, IEEE J. Sel. Top. Quantum Electron. \textbf{9}, 1502 (2003).






\bibitem{fuk11} D. Fukuda, G. Fujii, T. Numata, K. Amemiya, A. Yoshizawa, H. Tsuchida, H. Fujino, H. Ishii, T. Itatani, S. Inoue and T. Zama, \textit{Opt. Expr.} \textbf{19}, 870 (2011).

\bibitem{pre09} D. Prele, M.R. Piat, E.L. Breelle, F. Voisin, M. Pairat, Y. Atik, B. Belier, L. Dumoulin, C. Evesque, G. Klisnick, S. Marnieros, F. Pajot, M. Redon and G. Sou, \textit{IEEE T. Appl. Supercon.} \textbf{19}, 501 (2009).

\bibitem{cab08} B. Cabrera, \textit{J. Low Temp. Phys.} \textbf{151}, 82 (2008).

\bibitem{lit08} A.E. Lita, A.J. Miller and S.W. Nam, \textit{Opt. Expr.} \textbf{16}, 3032 (2008).

\bibitem{lol11} L. Lolli, G. Brida, I.P. Degiovanni, M. Gramegna, E. Monticone, F. Piacentini, C. Portesi, M. Rajteri, I. Ruo Berchera, E. Taralli, P. Traina, \textit{Int. J. of Quant. Inf.} \textbf{9}, 405 (2011).

\bibitem{lol13} L. Lolli, E. Taralli, C. Portesi, E. Monticone, and M. Rajteri, Appl. Phys. Lett. \textbf{103}, 041107 (2013)

\bibitem{ros05} D. Rosenberg, A.E. Lita, A.J. Miller and S.W. Nam, \textit{Phys. Rev. A} \textbf{71}, 061803 (2005).

\bibitem{bandler06} S.R. Bandler, E. Figueroa-Feliciano, N. Iyomoto, R.L. Kelley, C.A. Kilbourne, K.D. Murphy, F.S. Porter, T. Saab and J. Sadleir, \textit{Nucl. Instrum. Meth. A} \textbf{559}, 817 (2006).

\bibitem{irw05} K.D. Irwin and C.G. Hilton, in {\it Cryogenic Particle Detection}, C. Enss (Ed.), Topics Appl. Phys., \textbf{99} (Springer, Berlin, 2005).

\bibitem{Nam14} Z.H. Levine, B.L. Glebov, A.L. Migdall, T. Gerrits, B. Calkins, A.E. Lita and S.W. Nam, \textit{J. Opt. Soc. Am. B} \textbf{31}, B20 (2014).

\bibitem{Pfi14} N. Sridhar, R. Shahrokhshahi, A.J. Miller, B. Calkins, T. Gerrits, A. Lita, S.W. Nam, and O. Pfister, \textit{J. Opt. Soc. Am. B} \textbf{31}, B34 (2014).






\bibitem{multiSpatial} L.A. Jiang, E.A. Dauler and J.T. Chang, \textit{Phys. Rev. A} \textbf{75}, 062325 (2007).

\bibitem{multiSpatial1} A. Divochiy, F. Marsili, D. Bitauld, A. Gaggero, R. Leoni, F. Mattioli, A. Korneev, V. Seleznev, N. Kaurova, O. Minaeva, G. Goltsman, K.G. Lagoudakis, M. Benkhaoul, F. Le\`vy and A. Fiore, \textit{Nat. Phot.} \textbf{2}, 302 (2008).




\bibitem{multiTemporal1} K. Banaszek and I. A. Walmsley, Opt. Lett., \textbf{28} 52 (2003).




\bibitem{multiTemporal2} J. Řeháček, Z. Hradil, O. Haderka, J. Peřina, Jr., and M. Hamar, Phys. Rev. A 67, 061801(R) (2003).



\bibitem{multiTemporal3} D. Achilles, Ch. Silberhorn, C. Sliwa, K. Banaszek and I.A. Walmsley, \textit{Opt. Lett.} {\bf 28}, 2387 (2003).

\bibitem{multiTemporal4} M.J. Fitch, B.C. Jacobs, T.B. Pittman and J.D. Franson, \textit{Phys. Rev. A} \textbf{68}, 043814 (2003).



\bibitem{rad14} R. Chrapkiewicz, \textit{J. Opt. Soc. Am. B} \textbf{31}, B8 (2014).


\bibitem{gold13} E.A. Goldschmidt, F. Piacentini, I. Ruo Berchera, S.V. Polyakov, S. Peters, S. Kuck, G. Brida, I.P. Degiovanni, A. Migdall and M. Genovese, \textit{Phys. Rev. A} \textbf{88}, 013822 (2013).

\bibitem{but98} E.G. Atkinson and D.J. Butler, \textit{Metrologia} \textbf{35}, 241 (1998).


\bibitem{footnote} $\mathbf{x}=\{m \mathcal{C}, (4-m) \mathcal{N}\}$ corresponds for eaxample in the case $m=1$ to $\mathbf{x}= (\mathcal{C}, \mathcal{N}, \mathcal{N}, \mathcal{N})$, $(\mathcal{N}, \mathcal{C}, \mathcal{N}, \mathcal{N})$, $(\mathcal{N}, \mathcal{N}, \mathcal{C}, \mathcal{N})$, $(\mathcal{N}, \mathcal{N}, \mathcal{N}, \mathcal{C})$. Analogous arguments hold for the other cases.


\end{thebibliography}
\end{document}